\documentclass[sigconf]{acmart}

\AtBeginDocument{%
  \providecommand\BibTeX{{%
    \normalfont B\kern-0.5em{\scshape i\kern-0.25em b}\kern-0.8em\TeX}}}

\copyrightyear{2024}
\acmYear{2024}
\setcopyright{acmlicensed}
\acmConference[WWW '24 Companion] {Companion Proceedings of the ACM Web Conference 2024}{May 13--17, 2024}{Singapore, Singapore.}
\acmBooktitle{Companion Proceedings of the ACM Web Conference 2024 (WWW '24 Companion), May 13--17, 2024, Singapore, Singapore}
\acmISBN{979-8-4007-0172-6/24/05}
\acmDOI{10.1145/3589335.3651488}
\usepackage{fancyhdr}
\usepackage{amsmath,amsfonts}
\usepackage[linesnumbered,ruled,vlined]{algorithm2e}
\usepackage{pifont}
\usepackage{amsthm}
\usepackage{algpseudocode}
\usepackage{tcolorbox}
\usepackage{array}
\usepackage[caption=false,font=normalsize,labelfont=sf,textfont=sf]{subfig}
\usepackage{textcomp}
\usepackage{stfloats}
\usepackage{url}
\usepackage{verbatim}
\usepackage{graphicx}
\usepackage{ulem}
\usepackage{amsfonts}
\usepackage[utf8]{inputenc}
\usepackage{setspace}
\usepackage{cleveref}
\usepackage[utf8]{inputenc}
\crefname{section}{§}{§§}
\Crefname{section}{§}{§§}
\usepackage{caption}
\usepackage{booktabs}
\usepackage{makecell}
\usepackage{MnSymbol}
\usepackage{utfsym}
\usepackage{filecontents, pgffor}
\usepackage{balance}
\usepackage{latexsym}
\usepackage{color}

\usepackage{listings, xcolor}
\definecolor{verylightgray}{rgb}{.97,.97,.97}

\lstdefinelanguage{Solidity}{
	keywords=[1]{anonymous, assembly, assert, balance, break, call, callcode, case, catch, class, constant, continue, constructor, contract, debugger, default, delegatecall, delete, do, else, emit, event, experimental, export, external, false, finally, for, function, gas, if, implements, import, in, indexed, instanceof, interface, internal, is, length, library, log0, log1, log2, log3, log4, memory, modifier, new, payable, pragma, private, protected, public, pure, push, require, return, returns, revert, selfdestruct, send, solidity, storage, struct, suicide, super, switch, then, this, throw, transfer, true, try, typeof, using, value, view, while, with, addmod, ecrecover, keccak256, mulmod, ripemd160, sha256, sha3}, 
	keywordstyle=[1]\color{blue}\bfseries,
	keywords=[2]{address, bool, byte, bytes, bytes1, bytes2, bytes3, bytes4, bytes5, bytes6, bytes7, bytes8, bytes9, bytes10, bytes11, bytes12, bytes13, bytes14, bytes15, bytes16, bytes17, bytes18, bytes19, bytes20, bytes21, bytes22, bytes23, bytes24, bytes25, bytes26, bytes27, bytes28, bytes29, bytes30, bytes31, bytes32, enum, int, int8, int16, int24, int32, int40, int48, int56, int64, int72, int80, int88, int96, int104, int112, int120, int128, int136, int144, int152, int160, int168, int176, int184, int192, int200, int208, int216, int224, int232, int240, int248, int256, mapping, string, uint, uint8, uint16, uint24, uint32, uint40, uint48, uint56, uint64, uint72, uint80, uint88, uint96, uint104, uint112, uint120, uint128, uint136, uint144, uint152, uint160, uint168, uint176, uint184, uint192, uint200, uint208, uint216, uint224, uint232, uint240, uint248, uint256, var, void, ether, finney, szabo, wei, days, hours, minutes, seconds, weeks, years},	
	keywordstyle=[2]\color{teal}\bfseries,
	keywords=[3]{block, blockhash, coinbase, difficulty, gaslimit, number, timestamp, msg, data, gas, sender, sig, value, now, tx, gasprice, origin},	
	keywordstyle=[3]\color{violet}\bfseries,
	identifierstyle=\color{black},
	sensitive=true,
	comment=[l]{//},
	morecomment=[s]{/*}{*/},
	commentstyle=\color{gray}\ttfamily,
	stringstyle=\color{red}\ttfamily,
	morestring=[b]',
	morestring=[b]"
}

\begin{document}
\title{DeFiTail: DeFi Protocol Inspection through Cross-Contract Execution Analysis}
\author{Wenkai Li}

\affiliation{%
  \institution{Hainan University}
  \city{Haikou}
  \country{China}
}
\email{liwenkai871@gmail.com}

\author{Xiaoqi Li}
\authornote{The corresponding author}
\affiliation{%
  \institution{Hainan University}
  \city{Haikou}
  \country{China}
}
\email{csxqli@gmail.com}

\author{Yuqing Zhang}
\affiliation{%
  \institution{University of Chinese Academy of Sciences}
  \city{Beijing}
  \country{China}
}
\email{zhangyq@nipc.org.cn}

\author{Zongwei Li}
\affiliation{%
  \institution{Hainan University}
  \city{Haikou}
  \country{China}
}
\email{lizw1017@gmail.com}


\renewcommand{\shortauthors}{Wenkai Li, Xiaoqi Li, Yuqing Zhang, \& Zongwei Li}

\begin{abstract}
Decentralized finance (DeFi) protocols are crypto projects developed on the blockchain to manage digital assets. Attacks on DeFi have been frequent and have resulted in losses exceeding \$77 billion. However, detection methods for malicious DeFi events are still lacking. In this paper, we propose DeFiTail, the \textit{first} framework that utilizes deep learning to detect access control and flash loan exploits that may occur on DeFi. Since the DeFi protocol events involve invocations with multi-account transactions, which requires execution path unification with different contracts. Moreover, to mitigate the impact of mistakes in Control Flow Graph (CFG) connections, we validate the data path by employing the symbolic execution stack. Furthermore, we feed the data paths through our model to achieve the inspection of DeFi protocols. Experimental results indicate that DeFiTail achieves the highest accuracy, with 98.39\% in access control and 97.43\% in flash loan exploits. DeFiTail also demonstrates an enhanced capability to detect malicious contracts, identifying 86.67\% accuracy from the CVE dataset.
\end{abstract}


\maketitle

\section{Introduction}
\label{sec:intro}


Decentralized Finance (DeFi) protocols leverage smart contracts to build a payment ecosystem with financial transactions, such as Exchange, Lending, and Derivatives~\cite{zhang2022authros}.
Recent research \cite{RN66,RN65,li2022survey} has focused on detecting vulnerabilities in smart contracts, utilizing user-defined rules and expert knowledge to standardize detection capabilities. Moreover, the success of deep learning models has been proven in identifying correlation from historical contracts~\cite{chen2020finding,gao2019smartembed,li2023overview}, enabling them to discover vulnerable patterns that can capture fragile contracts. 

However, there are still some challenges in detecting DeFi projects. 
\textbf{Challenge 1 (C1): Invocation Pattern Learning.} Most execution processes in malicious DeFi events involve invocations of multiple contracts~\cite{REKT-database2023solidity}. However, the conventional approaches~\cite{gao2019smartembed,gao2020deep,zhuang2021smart,li2022security} focus on examining isolated data paths or execution paths within individual contracts, which is insufficient for inspecting DeFi protocols that contain lots of invocation patterns in multiple contracts.  
\textbf{Challenge 2 (C2): External and Internal Path Unification.} Data paths in DeFi protocols involve transaction-level external paths and code-level internal paths~\cite{zhang2020txspector}, where the external path means the execution flow between contracts, and the internal path means the control flow in a single contract. However, current studies~\cite{zhang2020txspector,wang2021blockeye, SmarTest2021UsenixSunbeom,ramezany2023midnight} focus on unifying contracts at the transaction level, lacking unifying external paths of DeFi protocols at the code level.
\textbf{Challenge 3 (C3): Data Path Feasibility Validation.} Previous studies~\cite{chen2019large,ma2021pluto} have revealed limitations in comprehending the data paths extracted from the Control Flow Graph (CFG). The smart contracts in a DeFi project can generate several CFGs, yielding multiple data paths after the CFGs connection. However, the different choices of entry points can determine different data paths extracted from the connected CFGs, generating different operational sequences. Therefore, it is essential to validate the feasibility of data paths at the operation level.

\textbf{Our Solution.} 
To address these challenges, we implement a tool called DeFiTail, which is a specialized DeFi inspection framework that learns the invocation patterns in data paths with deep learning. \textbf{Solution to C1}: We take advantage of sequence and graph learning technology, extracting sequential execution process features and structural heterogeneous graph features. First, we convert the data paths that contain opcodes and operands to a series of sequences according to the execution order in the connected CFGs. Then, a sequence learning model is used to learn the sequence features, and a graph learning model is used to build the heterogeneous graph for extracting the structural features. By combining the sequence and structural features, more complete features can be extracted.\\
\noindent\textbf{Solution to C2}: We analyze the external transactions in the Ethereum Virtual Machine (EVM), paralleling the internal transaction logic in the smart contract. Through the function segmentation with CFG construction, the 4-byte function signature in the caller function is obtained. Then, we assess whether an invocation exists between functions by examining the presence of the function signatures within another function, unifying external paths and internal paths.
\textbf{Solution to C3}: We integrate a symbolic execution stack into DeFiTail, enabling the validation of data path feasibility. The symbolic execution stack uses symbols to record the number of bytes in the stack. Comparing the equal relation between stack height and opcode rules in EVM can determine whether the path is feasible.

\textbf{Contributions.} The main contributions are as follows:
\begin{itemize}
    \item  We propose DeFiTail, to our best knowledge, which is the \textit{first} inspection framework with deep learning to detect DeFi attacks. It includes access control and flash loan exploits, on various EVM-compatible blockchains (\cref{sec::method}).
    \item We unify external and internal paths, and connect CFGs between bytecode contracts at the code level. Furthermore, the data path validation is formulated into the path reachability problem, identifying feasible data paths (\cref{subsec::dps}).
    \item We evaluate the DeFiTail outperformed SOTAs by 16.57\% and 11.26\% points in detecting access control and flash loan exploits. Moreover, we highlight the performance enhancements via CFG connection and data path validation (\cref{sec::evaluation}). 
    \item We open source DeFiTail at \cite{DeFiTail}, and further information will be accessible when the paper is accepted.
\end{itemize}

\section{DeFiTail}
\label{sec::method}

\begin{figure}[!t]
\centering
\includegraphics[width=0.9\linewidth]{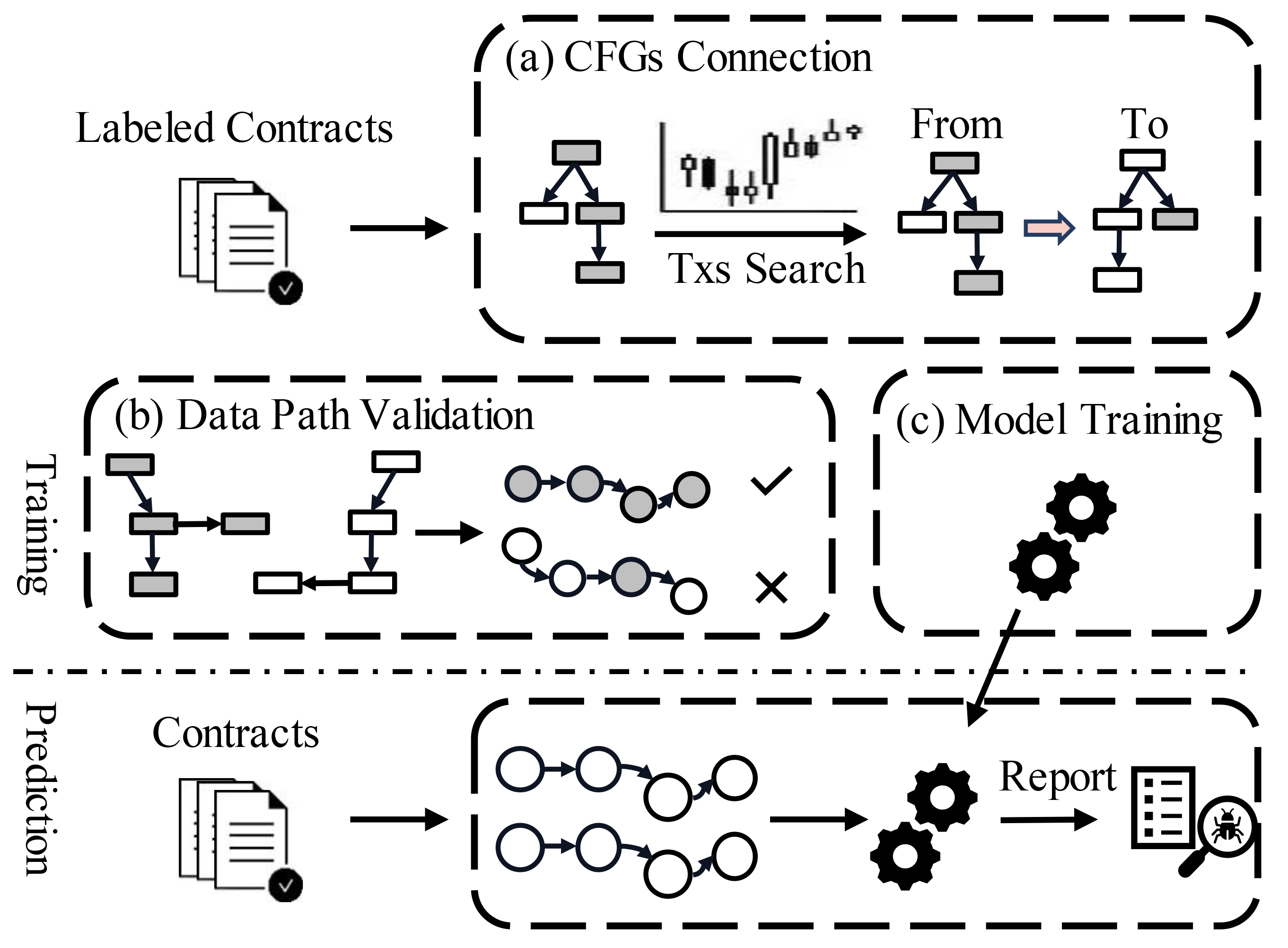}
\vspace{-3ex}
\caption{Overview of Framework. Above the dotted line is the training phase, and below the dotted line is the vulnerability prediction stage.}
\label{fig:framework}
\vspace{-4ex}
\end{figure}

As shown in Figure \ref{fig:framework}, our work comprises three steps. In the training phase, we feed the labeled DeFi protocols with contracts into DeFiTail, which outputs the bug report. Step (a) unifies the external and internal paths, and connects the CFGs (\cref{subsec::rcc}). Step (b) extracts the execution paths as data paths, and validates the data paths with the symbolic execution stack (\cref{subsec::dps}). Step (c) takes the valid data paths as input, training the model (\cref{subsec::model}). In the prediction phase, DeFiTail takes the bytecode contracts in DeFi protocols as input, and outputs the prediction results.

\subsection{CFGs Connection}
\label{subsec::rcc}

In the multi-contract call scenarios, the calling operations (i.e., \texttt{CALL}, \texttt{DELEGATECALL}, \texttt{STATICCALL}, \texttt{CALLCODE}) and the return operations (i.e., \texttt{RETURN}) are leveraged as foundational flags to delineate the control flows between contracts. As depicted in Figure \ref{fig:framework}, we categorize relevant contracts based on their functions and meticulously capture the function signature data for each constituent function throughout the process. When a function involves a calling operation, we divide the function into two segments. 


The segments are guided by the corresponding operations (e.g., \texttt{CALL}, \texttt{DELEGATECALL}, etc.). In the context of the \texttt{CALLDATA}, a function signature is located within the call data to which the calling operation is transitioned. For example,
the calling operations are recognized for caller function $\alpha$ within $CFG_t$. While the function interface of the callee function $\beta$ within $CFG_c$ matches the signature incorporated in the calldata of function $\alpha$. Then it proceeds to insert a control flow graph node, connecting different CFGs.

\subsection{Data Path Validation}
\label{subsec::dps}
When analyzing the connected CFG, the Depth-First Search (DFS) algorithm is utilized to select the data path that embeds as the contract feature. However, the path selected in this way may not constantly align with the real data flow. Since functions can be called at different entries, resulting in different data paths. For instance, the invocation of function $\mu$ at position $\hat{a}$ can occur both before and after the $\mu$ at $\hat{a}$. Therefore, to address the problem, a path validation analysis will be conducted to obtain a more precise data path feasibility.

\begin{figure}[ht]
\centering
\vspace{-2ex}
\includegraphics[width=0.95\linewidth]{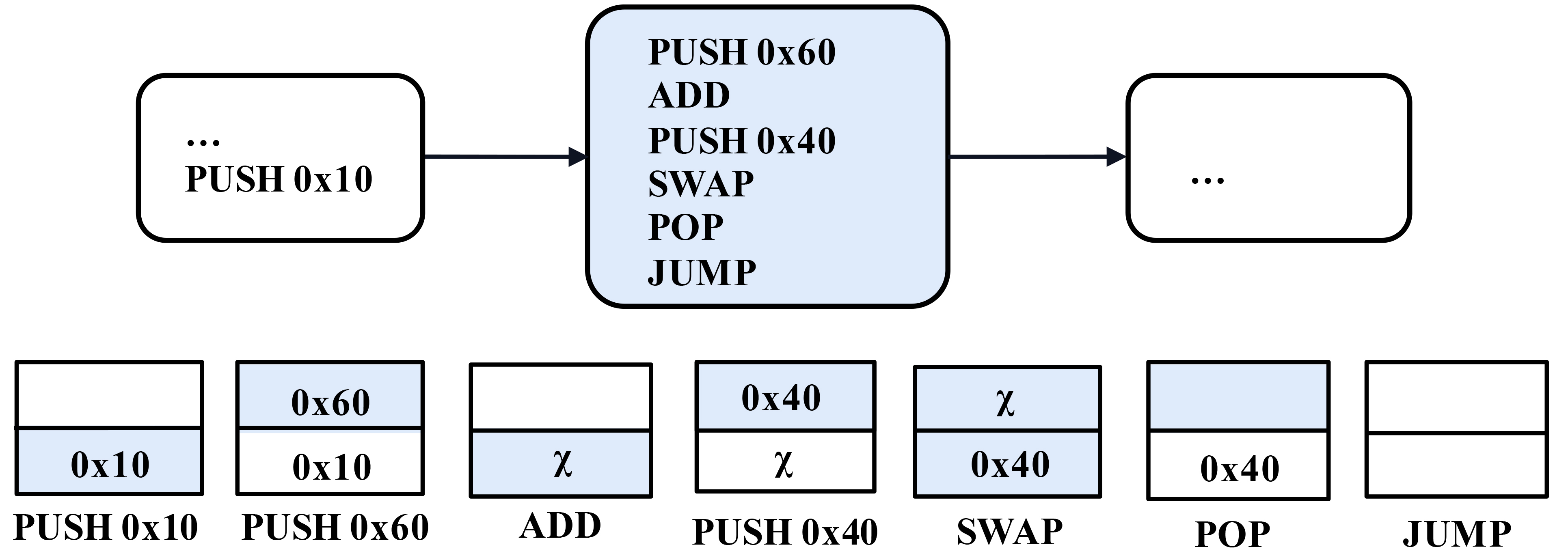}
\vspace{-2ex}
\caption{An Example of Symbolic Stack Execution. $\chi$ represents the placeholder of the calculation result.}
\label{fig:Symbolic_execution}
\vspace{-2ex}
\end{figure}


Figure \ref{fig:Symbolic_execution} depicts the working principle of the symbolic execution stack within the bytecode smart contracts. During this procedure, our primary focus is on the stack height, particularly the existence of a target signature during a jumping or calling operation. If the data path can fulfill the requirements of the symbolic execution stack comprehensively, the data path can be executed in an attainable manner, that is, the data path is feasible. Specifically, the function signature data includes various operations at the memory stack level, such as \texttt{PUSH}, \texttt{DUP}, \texttt{SWAP}, \texttt{AND}, and \texttt{POP}. The evaluation of calculation operation (e.g., \texttt{ADD}) results is circumvented by utilizing symbols as placeholders to maintain the stack height. 

\subsection{Model}
\label{subsec::model}
The model inputs all feasible data paths in the DeFi protocol that have removed the operands, then it outputs the classification results.

A data path embedding $DP_\pi$, which comprises a sequence of $n$ opcode embeddings represented as $\{t_1, t_2, ..., t_n\}$. Currently, sequence learning methods, such as the transformer, truncate each input to a fixed dimension, such as 512. However, the dimension is not enough to match the length of the data path that is generated by several contracts. Therefore, to suit DeFi contract scenarios, we establish a heterogeneous graph~\cite{zhuang2021smart} to learn the structural features that include the whole data path.

While constructing a graphical representation from the given data paths, we follow \cite{yao2019graph,lin2021bertgcn} to employ the data paths to create a node connection matrix, forming a heterogeneous graph. Initially, we construct a matrix $X^a$ with dimensions of $(n_{path}+n_{opcode})\times d$, wherein $n_{path}$ refers to the path node, $n_{opcode}$ pertains to the opcode node, and $d$ denotes the dimension of the feature embedding. The weight value is $PPMI(i,j)$ if the two nodes $i$ and $j$ are opcode nodes. In the scenario where $i$ is a path node and $j$ is an opcode node, the edge's weight value is deemed as $TF$-$IDF(i,j)$. If $i$ equals $j$, then we assign it a weight of 1. Otherwise, it is considered as 0. After completing all traversals, $X^a$ is converted into a triangular matrix $A_{ij}$ to hold the edge weights.

To learn the representation of path-path nodes on $X^a$, we utilize the 
sequence learning method to learn the path feature vector from the opcode sequence and then initialize it into the graph. In this step, we truncate the input path to a fixed length and convert it to a vector $X^b$ of dimensions ${(n_{path}+n_{opcode})\times d}$, aligning the heterogeneous graph vector. The corresponding dimension of opcode to 0, eliminating the influence of opcode features on graph nodes, i.e., $X^b=({X^a_{path}}^T, 0)^T$. Then, we select BERT as the transformer encoder to obtain the feature vector $f_{X^b}$ of each path $DP_\pi = \{\pi_1, \pi_2, ..., \pi_n\}$, which is used to initiate the path nodes in $X^a$. Simultaneously, we feed the path features to a linear and a softmax layer, demonstrated in Equation \ref{eqa:bert_classification}, to obtain the probability distribution. Note that we omit the bias to simply the explanation.

\vspace{-3ex}
\begin{equation}
   Y_{bert} = softmax(Wf_{X^b})
\label{eqa:bert_classification}
\end{equation}
\vspace{-2ex}

Once the path node initiation is completed on the heterogeneous graph $X^g$, we utilize a graph convolutional network (GCN) to extract features from all nodes. The GCN generates an output feature $X^{gcn}$ based on Equation \ref{eqa:GCN_matrix}. We then feed $X^{gcn}$ into a softmax layer to obtain the classification results.

\vspace{-3ex}
\begin{equation}
   X^{gcn} = \sigma(\widetilde{D}^{-\frac{1}{2}}\widetilde{A}\widetilde{D}^{-\frac{1}{2}}X^gW)
\label{eqa:GCN_matrix}
\end{equation}
\vspace{-2ex}
\begin{equation}
   Y_{gcn} = softmax(X^{gcn})
\label{eqa:output}
\end{equation}

The $\sigma$ is an activation function, the $\widetilde{D} = \sum_j\widetilde{A}_{ij}$ means the degree of the node $i$, and the $\widetilde{A}$ is an adjacency matrix. Thus, the $\widetilde{D}^{-\frac{1}{2}}\widetilde{A}\widetilde{D}^{-\frac{1}{2}}$ makes the $\widetilde{A}$ normalization. The $W \in \mathbb{R}^{C \times F}$ indicates a trainable parameters matrix, where $C$ expresses the dimension of the node feature vector, and $F$ represents the output dimension.


Subsequently, a linear interpolation is utilized to merge the inductive and transductive learning results, i.e., $Y = \lambda Y_{gcn} +(1-\lambda) Y_{bert}$. The $\lambda \in [0, 1]$ controls the balance of two probability distributions. 

\vspace{-2ex}
\begin{equation}
    loss(x,i) = W_i(-x_i +log(\sum_{j} exp(x_j)))
\label{eqa:crossentropy}
\end{equation}
\vspace{-2ex}

We simply employ the cross entropy loss function at Equation \ref{eqa:crossentropy} during the entire training, validation, and testing process. We also employ weight distributions $W=[W_1, W_2]$ for each category. The $W_i$ is computed as the total number of categories divided by the number of occurrences of the $i_{th}$ class. The non-softmax output is represented by $x=(x_0,x_1)$. Finally, we calculate the weighted loss for each class, then sum the weighted loss and divide it by the number of classes to get the final weighted-cross entropy loss.

\section{Experimental Evaluation}
\label{sec::evaluation}

In this section, we evaluate the performance of our tools and address the following questions: \underline{RQ1 (Accuracy):} Can DeFiTail outperform the state-of-the-art? \underline{RQ2 (Ablation):} How do distinct components impact the performance of DeFiTail? \underline{RQ3 (Applicability):}  Can DeFiTail identify vulnerabilities in the real world?

\textbf{Experimental Setup: }
we curated a dataset from the REKT database~\cite{REKT-database2023solidity}, which comprises a substantial collection of 14,301 data paths extracted from 3,216 instances of hacked DeFi events. Significantly, these malicious activities transpired across EVM-compatible blockchains, including Ethereum, Binance, Arbitrum, Fantom, etc. In the RQ1 and RQ2, we randomly select 80\% for training, and the remaining 20\% for testing. We use the Accuracy to evaluate the performance since the dataset cannot fairly judge the true or false positive of the result. In the Transformer encoder, the embedding size is 128, the number of layers is 4, the hidden size is 312, the feed-forward size is 1200, the head number is 12, and the learning rate is 1e-5. In the GCN, the number of layers is 2, the hidden size is 200, and the learning rate is 1e-3. The optimizer is Adam. The dropout is set to 0.5, and the batch size is 64. All experiments are conducted on a server with an NVIDIA GeForce GTX 4070TI GPU and an Intel(R) Core(TM) i9-13900KF CPU.

\textbf{RQ1 Accuracy:}
\label{subsec::comparison}
We train DeFiTail with the REKT dataset~\cite{REKT-database2023solidity} and test it with the state-of-the-art, whose comparative results are shown in Table \ref{tab:comparison_results_interaction}. As for access control, we compare DeFiTail with Mythril~\cite{mythril2019}, Ethainter~\cite{brent2020ethainter}, and Achecker~\cite{ghaleb2023achecker}. We follow the definition of access control in Achecker~\cite{ghaleb2023achecker}, the SWC-\{105,106,136,109\} in Mythril and Ethainter are judged as access control events. When using Mythril, specific related vulnerabilities are deemed relevant as access control is not directly detected. In the situations of real DeFi scams, the accuracy in the table will be lower than 64.91\%. Since the Ethainter analyzes contracts on three different blockchains, we detected all available contracts and calculated an accuracy of 42.86\%. Note that the Ethainter provides a public website\footnote{https://library.dedaub.com/}, and the comparison results in Table \ref{tab:comparison_results_interaction} are from it. The Achecker is a tool that is commonly utilized to identify permission vulnerabilities. After testing, Achecker achieved an accuracy of 81.82\%. DeFiTail achieved 98.39\% accuracy in the test set, outperforming other SOTA tools in Table \ref{tab:comparison_results_interaction}. As for flash loan exploits, there is limited research on it~\cite{wang2021towards,qin2021attacking,EEAtools2023quantstamp,ramezany2023midnight}, and the only open-source tool we found is Midnight~\cite{ramezany2023midnight}, which is a real-time tool based on event analysis. Midnight associates events with contracts based on real-time transactions, so we feed the transaction records when the events happended in the REKT database. Finally, we compare the results of DeFiTail and Midnight in Table \ref{tab:comparison_results_interaction}. \textbf{Answer to RQ1: } DeFiTail exhibits remarkable efficiency in the detection of access control and flash loan exploits, achieving accuracies of 98.39\% and 97.43\%, respectively.

\begin{table}[ht]
\footnotesize
\centering
  \caption{The Comparison Results in Different Tools}
  \vspace{-2ex}
  \label{tab:comparison_results_interaction}
  \begin{tabular}{c c c }
    \toprule
    \textbf{Model Name} & \textbf{Access Control }& \textbf{Flash Loan Exploits} \\
    \midrule
    Mythril\cite{mythril2019} & 64.91\% & --\\
    Ethainter\cite{brent2020ethainter} & 42.86\% &--\\
    AChecker\cite{ghaleb2023achecker} &81.82\% & --\\
    Midnight\cite{ramezany2023midnight} &-- & 86.17\% \\
    DeFiTail & 98.39\%& 97.43\%\\ 
  \bottomrule
  \vspace{-5ex}
\end{tabular}
\end{table}

\textbf{RQ2 Ablation:}
We evaluate the individual impact of CFGs connection and data path validation, and the results are presented in Table \ref{tab:ablation_analysis}. Furthermore, the effects of the Transformer and the heterogeneous graph are investigated. 
As for the access control, DeFiTail performs 71.65\% and 0.16\% better with CFGs connection and data path validation than without them. The detection accuracy of access control is surprisingly improved by data path validation. However, the detection performance experiences a significant decline in the absence of CFGs connection. 
As for flash loan events, the results are reversed. The detection of flash loan events is negatively affected by only CFGs connection without the data path validation process. DeFiTail-np only includes the CFGs connection, whose accuracy is significantly lower at 12.57\% compared to DeFiTail. Due to the minor difference of only 0.21\% point between DeFiTail and DeFiTail-npc, which lacks both features, it suggests that the CFGs connection will enhance the detection results. Therefore, it is evident that the data path verification when CFGs connection has a positive impact on the detection results. Additionally, we compare the methods that directly utilize sequence models, including LSTM and BERT, in Table \ref{tab:ablation_analysis}. The results show that access control detection is influenced by different components (LSTM, BERT, and GCN), which significantly impact access control detection. However, there is a slight discrepancy in the effectiveness of the three detection techniques in identifying flash loan exploits. \textbf{Answer to RQ2: } The presence of both path selection and related CFG connection in access control and flash loan exploits is crucial for the dependencies of DeFiTail.


\begin{table}[ht]
\footnotesize
\centering
\vspace*{-2ex}
  \caption{The Ablation Analysis Results. np/npc means the model without path validation/CFGs connection. BERT represents the model with the pre-trained BERT-base.}
  \vspace{-2ex}
  \label{tab:ablation_analysis}
  \begin{tabular}{c c c }
    \toprule
    \textbf{Model Name} & \textbf{Access Control} & \textbf{Flash Loan Exploits}\\
    \midrule
    DeFiTail-np         & 98.23\%   & 85.19\%  \\
    DeFiTail-npc        & 57.33\%   & 97.22\%  \\
    \hline
    DeFiTail-LSTM       & 88.89\%   & 95.00\%  \\
    DeFiTail-BERT       & 89.71\%   & 96.30\%  \\
    \textbf{DeFiTail}   & 98.39\%   & 97.43\%  \\
  \bottomrule
\end{tabular}
\vspace{-2ex}
\end{table}

\textbf{RQ3 Applicability:}
\label{subsec::case}
To verify the effectiveness of DeFiTail in inspecting access control events in the real environment, we detected CVE vulnerabilities listed in Table \ref{tab:case_analysis} with Mythril, SPCon~\cite{liu2022finding}, and DeFiTail. The CVE dataset has been checked by the authors of SPCon and AChecker~\cite{ghaleb2023achecker}. Table \ref{tab:case_analysis} indicates that the tag "N/A" denotes inadequate input data necessary for analysis. For example, CVE-2018-19830, CVE-2018-19833, and CVE-2018-19834 are all designated as "N/A" on SPCon because of insufficient transaction data for historical role mining simulations related to access control or permission issues. Additionally, CVE-2018-10705 is marked "N/A" in the DeFiTail because there are not enough contracts in the malicious events. Moreover, to ensure maximum detection, we performed vulnerability events using Mythril for at least 30 minutes on the cases analyzed in table \ref{tab:case_analysis}. \textbf{Answer to RQ3: } DeFiTail exhibited a remarkable ability to identify security vulnerabilities, successfully detecting 86.67\% of 15 CVE incidents.

\begin{table}[h]
\footnotesize
\centering
  \caption{The Evaluation Results on CVE Dataset. N/A means that necessary information is missing.}
  \vspace{-2ex}
  \label{tab:case_analysis}
  \begin{tabular}{c c c c c c}
    \toprule
    \textbf{ID} & \textbf{CVE-ID} & \textbf{Mythril} & \textbf{SPCon} & \textbf{AChecker}  & \textbf{DeFiTail}\\
    \midrule
    1 &  CVE-2018-10666  & \scalebox{0.75}{\usym{2613}}  & $\checkmark$  & $\checkmark$  & $\checkmark$  \\
    2 &  CVE-2018-10705  & \scalebox{0.75}{\usym{2613}}  & $\checkmark$  & $\checkmark$  & N/A  \\
    3 &  CVE-2018-11329  & \scalebox{0.75}{\usym{2613}}  & $\checkmark$  & $\checkmark$ & \scalebox{0.75}{\usym{2613}} \\
    4 &  CVE-2018-19830  & $\checkmark$   & N/A & $\checkmark$ & $\checkmark$  \\
    5 &  CVE-2018-19831  & \scalebox{0.75}{\usym{2613}}  & $\checkmark$  & $\checkmark$  & $\checkmark$  \\
    6 &  CVE-2018-19832  & \scalebox{0.75}{\usym{2613}}  & $\checkmark$ & $\checkmark$  & $\checkmark$  \\
    7 &  CVE-2018-19833  & \scalebox{0.75}{\usym{2613}}  & N/A & $\checkmark$ & $\checkmark$  \\
    8 &  CVE-2018-19834  & \scalebox{0.75}{\usym{2613}}  & N/A & $\checkmark$ & $\checkmark$  \\
    9 &  CVE-2019-15078  & $\checkmark$  & $\checkmark$ & $\checkmark$  & $\checkmark$  \\
    10 &  CVE-2019-15079  & \scalebox{0.75}{\usym{2613}}  & $\checkmark$ & \scalebox{0.75}{\usym{2613}} & $\checkmark$  \\
    11 &  CVE-2019-15080  & \scalebox{0.75}{\usym{2613}}  & $\checkmark$ & $\checkmark$  & $\checkmark$ \\
    12 &  CVE-2020-17753  & $\checkmark$  & \scalebox{0.75}{\usym{2613}} & \scalebox{0.75}{\usym{2613}} & $\checkmark$  \\
    13 &  CVE-2020-35962  & $\checkmark$   & \scalebox{0.75}{\usym{2613}} & \scalebox{0.75}{\usym{2613}} & $\checkmark$  \\
    14 &  CVE-2021-34272  & \scalebox{0.75}{\usym{2613}}  & $\checkmark$ & $\checkmark$ & $\checkmark$  \\
    15 &  CVE-2021-34273  & \scalebox{0.75}{\usym{2613}}  & \scalebox{0.75}{\usym{2613}} & $\checkmark$  & $\checkmark$  \\
  \bottomrule
  \vspace{-7ex}
\end{tabular}
\end{table}

\section{Conclusion}
\label{sec::conclusion}
\vspace{-1ex}
The paper proposes DeFiTail, a new framework for detecting DeFi protocols at the bytecode level. To capture the invocation patterns of DeFi protocols, we extract the data paths, and then use sequence and graph learning to learn the patterns. To validate the correctness of CFG, we rely on the symbolic execution stack to verify the path feasibility of the connected CFG. Experiments reveal that DeFiTail outperforms the SOTA in access control and flash loan exploits.

\vspace{-1ex}
\section{ACKNOWLEDGMENTS}
\vspace{-1ex}
This work is sponsored by the National Natural Science Foundation of China (No.62362021), CCF-Tencent Rhino-Bird Open Research Fund (No.RAGR20230115), and Hainan Provincial Department of Education Project (No.HNJG2023-10).
\vspace{-1ex}
\normalem
\bibliographystyle{ACM-Reference-Format}
\vspace{-1ex}
\bibliography{main}


\begin{thebibliography}{27}


\ifx \showCODEN    \undefined \def \showCODEN     #1{\unskip}     \fi
\ifx \showDOI      \undefined \def \showDOI       #1{#1}\fi
\ifx \showISBNx    \undefined \def \showISBNx     #1{\unskip}     \fi
\ifx \showISBNxiii \undefined \def \showISBNxiii  #1{\unskip}     \fi
\ifx \showISSN     \undefined \def \showISSN      #1{\unskip}     \fi
\ifx \showLCCN     \undefined \def \showLCCN      #1{\unskip}     \fi
\ifx \shownote     \undefined \def \shownote      #1{#1}          \fi
\ifx \showarticletitle \undefined \def \showarticletitle #1{#1}   \fi
\ifx \showURL      \undefined \def \showURL       {\relax}        \fi
\providecommand\bibfield[2]{#2}
\providecommand\bibinfo[2]{#2}
\providecommand\natexlab[1]{#1}
\providecommand\showeprint[2][]{arXiv:#2}

\bibitem[Brent et~al\mbox{.}(2020)]%
        {brent2020ethainter}
\bibfield{author}{\bibinfo{person}{Lexi Brent}, \bibinfo{person}{Neville Grech}, \bibinfo{person}{Sifis Lagouvardos}, {and} \bibinfo{person}{et al.}}
\newblock \showarticletitle{Ethainter: a smart contract security analyzer for composite vulnerabilities} \bibinfo{year}{2020}\natexlab{}. In \bibinfo{booktitle}{\emph{Proc. of PLDI}}. \bibinfo{pages}{454--469}.
\newblock


\bibitem[Chen(2020)]%
        {chen2020finding}
\bibfield{author}{\bibinfo{person}{Jiachi Chen}.}
\newblock \showarticletitle{Finding Ethereum Smart Contracts Security Issues by Comparing History Versions} \bibinfo{year}{2020}\natexlab{}. In \bibinfo{booktitle}{\emph{Proc. of ASE}}. \bibinfo{pages}{1382--1384}.
\newblock


\bibitem[Chen et~al\mbox{.}(2019)]%
        {chen2019large}
\bibfield{author}{\bibinfo{person}{Ting Chen}, \bibinfo{person}{Zihao Li}, \bibinfo{person}{Yufei Zhang}, {and} \bibinfo{person}{et al.}}
\newblock \showarticletitle{A large-scale empirical study on control flow identification of smart contracts} \bibinfo{year}{2019}\natexlab{}. In \bibinfo{booktitle}{\emph{Proc. of ESEM}}. \bibinfo{pages}{1--11}.
\newblock


\bibitem[Engineers(2023)]%
        {REKT-database2023solidity}
\bibfield{author}{\bibinfo{person}{Solidity Engineers}.}
\newblock \bibinfo{title}{Top Crypto Hacks} \bibinfo{year}{2023}\natexlab{}.
\newblock
\newblock
\urldef\tempurl%
\url{https://de.fi/rekt-database}
\showURL{%
\tempurl}


\bibitem[Gao(2020)]%
        {gao2020deep}
\bibfield{author}{\bibinfo{person}{Zhipeng Gao}.}
\newblock \showarticletitle{When deep learning meets smart contracts} \bibinfo{year}{2020}\natexlab{}. In \bibinfo{booktitle}{\emph{Proc. of ASE}}. \bibinfo{pages}{1400--1402}.
\newblock


\bibitem[Gao et~al\mbox{.}(2019)]%
        {gao2019smartembed}
\bibfield{author}{\bibinfo{person}{Zhipeng Gao}, \bibinfo{person}{Vinoj Jayasundara}, \bibinfo{person}{Lingxiao Jiang}, \bibinfo{person}{Xin Xia}, \bibinfo{person}{David Lo}, {and} \bibinfo{person}{John Grundy}.}
\newblock \showarticletitle{Smartembed: A tool for clone and bug detection in smart contracts through structural code embedding} \bibinfo{year}{2019}\natexlab{}. In \bibinfo{booktitle}{\emph{Proc. of ICSME}}. \bibinfo{pages}{394--397}.
\newblock


\bibitem[Ghaleb et~al\mbox{.}(2023)]%
        {ghaleb2023achecker}
\bibfield{author}{\bibinfo{person}{Asem Ghaleb}, \bibinfo{person}{Julia Rubin}, {and} \bibinfo{person}{et al.}}
\newblock \showarticletitle{AChecker: Statically Detecting Smart Contract Access Control Vulnerabilities} \bibinfo{year}{2023}\natexlab{}. In \bibinfo{booktitle}{\emph{Proc. of ICSE}}. \bibinfo{pages}{1--12}.
\newblock


\bibitem[Li(2023)]%
        {DeFiTail}
\bibfield{author}{\bibinfo{person}{Wenkai Li}.}
\newblock \bibinfo{title}{DeFiTail artifact} \bibinfo{year}{2023}\natexlab{}.
\newblock
\newblock
\newblock
\shownote{\url{http://doi.org/10.6084/m9.figshare.24117993}}.


\bibitem[Li et~al\mbox{.}(2022a)]%
        {li2022security}
\bibfield{author}{\bibinfo{person}{Wenkai Li}, \bibinfo{person}{Jiuyang Bu}, \bibinfo{person}{Xiaoqi Li}, {and} \bibinfo{person}{Xianyi Chen}.}
\newblock \showarticletitle{Security analysis of DeFi: Vulnerabilities, attacks and advances} \bibinfo{year}{2022}\natexlab{a}. In \bibinfo{booktitle}{\emph{Proc. of Blockchain}}. \bibinfo{pages}{488--493}.
\newblock


\bibitem[Li et~al\mbox{.}(2022b)]%
        {li2022survey}
\bibfield{author}{\bibinfo{person}{Wenkai Li}, \bibinfo{person}{Jiuyang Bu}, \bibinfo{person}{Xiaoqi Li}, \bibinfo{person}{Hongli Peng}, \bibinfo{person}{Yuanzheng Niu}, {and} \bibinfo{person}{Yuqing Zhang}.}
\newblock \showarticletitle{A survey of DeFi security: Challenges and opportunities} \bibinfo{year}{2022}\natexlab{b}.
\newblock \bibinfo{journal}{\emph{Journal of King Saud University-Computer and Information Sciences}} \bibinfo{volume}{34}, \bibinfo{number}{10}, \bibinfo{pages}{10378--10404}.
\newblock


\bibitem[Li et~al\mbox{.}(2021)]%
        {RN65}
\bibfield{author}{\bibinfo{person}{Xiaoqi Li}, \bibinfo{person}{Ting Chen}, \bibinfo{person}{Xiapu Luo}, {and} \bibinfo{person}{Chenxu Wang}.}
\newblock \showarticletitle{CLUE: towards discovering locked cryptocurrencies in ethereum} \bibinfo{year}{2021}\natexlab{}. In \bibinfo{booktitle}{\emph{Proc. of SAC}}. \bibinfo{pages}{1584--1587}.
\newblock


\bibitem[Li et~al\mbox{.}(2020)]%
        {RN66}
\bibfield{author}{\bibinfo{person}{Xiaoqi Li}, \bibinfo{person}{Ting Chen}, \bibinfo{person}{Xiapu Luo}, {and} \bibinfo{person}{Jiangshan Yu}.}
\newblock \showarticletitle{Characterizing erasable accounts in ethereum} \bibinfo{year}{2020}\natexlab{}. In \bibinfo{booktitle}{\emph{Proc. of ISC}}. \bibinfo{pages}{352--371}.
\newblock


\bibitem[Li et~al\mbox{.}(2023)]%
        {li2023overview}
\bibfield{author}{\bibinfo{person}{Zongwei Li}, \bibinfo{person}{Dechao Kong}, \bibinfo{person}{Yuanzheng Niu}, \bibinfo{person}{Hongli Peng}, \bibinfo{person}{Xiaoqi Li}, {and} \bibinfo{person}{Wenkai Li}.}
\newblock \showarticletitle{An Overview of AI and Blockchain Integration for Privacy-Preserving} \bibinfo{year}{2023}\natexlab{}.
\newblock \bibinfo{journal}{\emph{arXiv preprint arXiv:2305.03928}}.
\newblock


\bibitem[Lin et~al\mbox{.}(2021)]%
        {lin2021bertgcn}
\bibfield{author}{\bibinfo{person}{Yuxiao Lin}, \bibinfo{person}{Yuxian Meng}, \bibinfo{person}{Xiaofei Sun}, {and} \bibinfo{person}{et al.}}
\newblock \showarticletitle{BertGCN: Transductive Text Classification by Combining GNN and BERT} \bibinfo{year}{2021}\natexlab{}. In \bibinfo{booktitle}{\emph{Proc. of ACL}}. \bibinfo{pages}{1456--1462}.
\newblock


\bibitem[Liu et~al\mbox{.}(2022)]%
        {liu2022finding}
\bibfield{author}{\bibinfo{person}{Ye Liu}, \bibinfo{person}{Yi Li}, \bibinfo{person}{Shang-Wei Lin}, {and} \bibinfo{person}{Cyrille Artho}.}
\newblock \showarticletitle{Finding permission bugs in smart contracts with role mining} \bibinfo{year}{2022}\natexlab{}. In \bibinfo{booktitle}{\emph{Proc. of ISSTA}}. \bibinfo{pages}{716--727}.
\newblock


\bibitem[Ma and et~al.(2021)]%
        {ma2021pluto}
\bibfield{author}{\bibinfo{person}{Fuchen Ma} {and} \bibinfo{person}{et al.}}
\newblock \showarticletitle{Pluto: Exposing vulnerabilities in inter-contract scenarios} \bibinfo{year}{2021}\natexlab{}.
\newblock \bibinfo{journal}{\emph{IEEE Transactions on Software Engineering}} \bibinfo{volume}{48}, \bibinfo{number}{11}, \bibinfo{pages}{4380--4396}.
\newblock


\bibitem[Mythril(2019)]%
        {mythril2019}
\bibfield{author}{\bibinfo{person}{Mythril}.}
\newblock \bibinfo{title}{A Security Analysis Tool.} \bibinfo{year}{2019}\natexlab{}.
\newblock
\newblock
\urldef\tempurl%
\url{https://github.com/ConsenSys/mythril}
\showURL{%
\tempurl}


\bibitem[Qin et~al\mbox{.}(2021)]%
        {qin2021attacking}
\bibfield{author}{\bibinfo{person}{Kaihua Qin}, \bibinfo{person}{Liyi Zhou}, \bibinfo{person}{Benjamin Livshits}, {and} \bibinfo{person}{Arthur Gervais}.}
\newblock \showarticletitle{Attacking the defi ecosystem with flash loans for fun and profit} \bibinfo{year}{2021}\natexlab{}. In \bibinfo{booktitle}{\emph{Proc. of FC}}. \bibinfo{pages}{3--32}.
\newblock


\bibitem[Quantstamp(2023)]%
        {EEAtools2023quantstamp}
\bibfield{author}{\bibinfo{person}{Quantstamp}.}
\newblock \bibinfo{title}{Quantstamp} \bibinfo{year}{2023}\natexlab{}.
\newblock
\newblock
\urldef\tempurl%
\url{https://quantstamp.com/economic-exploits}
\showURL{%
\tempurl}


\bibitem[Ramezany et~al\mbox{.}(2023)]%
        {ramezany2023midnight}
\bibfield{author}{\bibinfo{person}{Shahin Ramezany}, \bibinfo{person}{Rachsuda Setthawong}, {and} \bibinfo{person}{Pisal Setthawong}.}
\newblock \showarticletitle{Midnight: An Efficient Event-driven EVM Transaction Security Monitoring Approach For Flash Loan Detection} \bibinfo{year}{2023}\natexlab{}. In \bibinfo{booktitle}{\emph{Proc. of JCSSE}}. \bibinfo{pages}{237--241}.
\newblock


\bibitem[So et~al\mbox{.}(2021)]%
        {SmarTest2021UsenixSunbeom}
\bibfield{author}{\bibinfo{person}{Sunbeom So}, \bibinfo{person}{Seongjoon Hong}, {and} \bibinfo{person}{Hakjoo Oh}.}
\newblock \showarticletitle{SmarTest: Effectively Hunting Vulnerable Transaction Sequences in Smart Contracts through Language Model-Guided Symbolic Execution} \bibinfo{year}{2021}\natexlab{}. In \bibinfo{booktitle}{\emph{Proc. of USENIX Security}}. \bibinfo{pages}{17--20}.
\newblock


\bibitem[Wang et~al\mbox{.}(2021a)]%
        {wang2021blockeye}
\bibfield{author}{\bibinfo{person}{Bin Wang}, \bibinfo{person}{Han Liu}, \bibinfo{person}{Chao Liu}, \bibinfo{person}{Zhiqiang Yang}, \bibinfo{person}{Qian Ren}, {and} \bibinfo{person}{et al.}}
\newblock \showarticletitle{Blockeye: Hunting for defi attacks on blockchain} \bibinfo{year}{2021}\natexlab{a}. In \bibinfo{booktitle}{\emph{Proc. of ICSE}}. \bibinfo{pages}{17--20}.
\newblock


\bibitem[Wang et~al\mbox{.}(2021b)]%
        {wang2021towards}
\bibfield{author}{\bibinfo{person}{Dabao Wang}, \bibinfo{person}{Siwei Wu}, {and} \bibinfo{person}{et al.}}
\newblock \showarticletitle{Towards a first step to understand flash loan and its applications in defi ecosystem} \bibinfo{year}{2021}\natexlab{b}. In \bibinfo{booktitle}{\emph{Proc. of ASIA-CCS}}. \bibinfo{pages}{23--28}.
\newblock


\bibitem[Yao et~al\mbox{.}(2019)]%
        {yao2019graph}
\bibfield{author}{\bibinfo{person}{Liang Yao}, \bibinfo{person}{Chengsheng Mao}, {and} \bibinfo{person}{Yuan Luo}.}
\newblock \showarticletitle{Graph convolutional networks for text classification} \bibinfo{year}{2019}\natexlab{}. In \bibinfo{booktitle}{\emph{Proc. of AAAI}}. \bibinfo{pages}{7370--7377}.
\newblock


\bibitem[Zhang et~al\mbox{.}(2020)]%
        {zhang2020txspector}
\bibfield{author}{\bibinfo{person}{Mengya Zhang}, \bibinfo{person}{Xiaokuan Zhang}, {and} \bibinfo{person}{et al.}}
\newblock \showarticletitle{TXSPECTOR: Uncovering attacks in ethereum from transactions} \bibinfo{year}{2020}\natexlab{}. In \bibinfo{booktitle}{\emph{Proc. of USENIX Security}}. \bibinfo{pages}{2775--2792}.
\newblock


\bibitem[Zhang et~al\mbox{.}(2022)]%
        {zhang2022authros}
\bibfield{author}{\bibinfo{person}{Shenhui Zhang}, \bibinfo{person}{Wenkai Li}, \bibinfo{person}{Xiaoqi Li}, {and} \bibinfo{person}{Boyi Liu}.}
\newblock \showarticletitle{Authros: Secure data sharing among robot operating systems based on ethereum} \bibinfo{year}{2022}\natexlab{}. In \bibinfo{booktitle}{\emph{Proc. of QRS}}. \bibinfo{pages}{147--156}.
\newblock


\bibitem[Zhuang et~al\mbox{.}(2021)]%
        {zhuang2021smart}
\bibfield{author}{\bibinfo{person}{Yuan Zhuang}, \bibinfo{person}{Zhenguang Liu}, \bibinfo{person}{Peng Qian}, {and} \bibinfo{person}{et al.}}
\newblock \showarticletitle{Smart contract vulnerability detection using graph neural networks} \bibinfo{year}{2021}\natexlab{}. In \bibinfo{booktitle}{\emph{Proc. of IJCAI}}. \bibinfo{pages}{3283--3290}.
\newblock


\end{thebibliography}

\end{document}